\documentclass[preprint]{aastex}        

\usepackage{amsmath}                     
\usepackage{natbib}  			
\usepackage{pbox}				
\usepackage{enumerate}               

\newcommand{\citeinpar}[1]{\citeauthor{#1} \citeyear{#1}}

\begin{document}

\title{Mentoring partnerships in science education}
\author{Andria C. Schwortz$^{a, b}$}
\author{Andrea C. Burrows$^c$}
\author{Sarah Katie Guffey$^c$}
\affil{$^a$Physics \& Astronomy, University of Wyoming, Laramie, WY, USA}
\affil{$^b$Natural Sciences, Quinsigamond Community College, Worcester, MA, USA}
\affil{$^c$Secondary Education, University of Wyoming, Laramie, WY, USA}
\email{aschwort@uwyo.edu}


\begin{abstract}
The authors use an action research (AR) approach in a collegiate studio physics class to investigate the power of partnerships via conferences as they relate to issues of establishing a student/mentor rapport, empowering students to reduce inequity, and the successes and barriers to hearing students' voices. 
The graduate teaching assistant (TA, Author 1) conducted one-on-one conferences with 29 students, elicited student opinions about the progress of the course, and talked with faculty, TAs, and an undergraduate supplemental instructor for other sections of the course. 
At the end of the semester, the students reported increased knowledge of the TA as a person and as an instructor, and vice versa. 
Sixty-five percent of students reported no interest in changing circumstances to make it easier to talk about personal concerns with the TA. 
College students reluctantly voiced their opinions about the course, possibly due to the power structure of the classroom. 
Other TAs in the department expressed mostly disinterest in the project, while faculty members were interested in student learning but skeptical of student empowerment. 
A case study of one student is presented, wherein his attendance improved in the course and he received additional help outside class, both possibly as a result of the student/TA conferences. 
Students in this studio physics section were more likely to interact directly with faculty or TAs during lectures, but less likely to do so during lab sessions, than were students in a non-studio physics section.
\end{abstract}


\section{Introduction}\label{sec:Intro}
Teaching at all levels and subjects requires a relationship between instructor and students (for example, \citeinpar{B11}). 
Instructors grow to know their students' academic strengths and weaknesses through pedagogical content knowledge \citep{Kriner}.
Students must trust their instructors to guide them through their challenges, even when unaware of these challenges. 
A partnership requires give and take between both sides of the educational equation, with both sides providing feedback to the other as they work towards the mutual goal of student success \citep{B11}. 
Yet so much of education, in K-12 and especially in college, continues to follow the traditional model, where the instructor is the source of all knowledge and the student's job is to passively absorb information (for example, \citeinpar{Breslow}).

The authors use action research (AR) in a collegiate studio physics classroom in order to break down the traditional academic model. All three authors are experienced science, technology, engineering, and math (STEM) educators at the secondary and college level and came to this project with questions about improving their teaching within an AR perspective. They saw multiple instances of inequity among collegiate students despite the equal access to education afforded by the public system. For example, while US K-12 students might have the right to secondary education, accessing education can be difficult when they are responsible for helping to raise siblings while their parents work multiple jobs. All students in both secondary and college classrooms may be able or even required to take science courses; however, these science courses may be viewed as just another pointless hurdle in schooling. 
`Equal access' to a resource, such as education, is not sufficient in the context of complex overlapping inequities of race, gender, income, family educational history, current family status, geography, and more. The authors address some of this classroom inequity by providing a platform for a larger student voice and by supporting them in establishing a rapport with one or more mentors in science.

This AR study is about mentoring partnerships in collegiate science education. The influence of conferences on college physics students with a graduate teaching assistant (TA) is important, as there are a number of current gaps in the literature. The gaps include a dearth of qualitative studies on science classroom environments, student/teacher conferences in science, and approaches to reducing the domination of the classroom by the teacher. Accordingly, `The challenge for all of us is to continue to think about how the curriculum can be used to improve the teaching and learning in science for all' (\citeinpar{DeBoer}, 576). The curriculum in this study is more than just the content, but includes the relationship between the student and the TA.

This study was set in a collegiate calculus-based Physics II course for physics majors and engineers at a research university in the western United States. The course contained 36 students, and all were engineering majors or engineering double-majors. Physics II is traditionally taught in a teacher-centered format with separate lecture, lab, and discussion sections. A faculty member typically teaches the lecture, whereas different TAs instruct the lab and discussion sections, all of which are typically held in different locations and taught by different methods. For example, lectures may be focused on conceptual ideas and discussion on problem-solving. The authors studied a section taught in a studio physics format without distinction between the different modalities of lecture, lab, and discussion. The faculty member and TA (Author 1) were present for nearly all of the sessions.

Physics is traditionally taught via a teacher-centered format. Studio physics is an attempt at changing this power structure (as discussed in the literature Review section). The authors explored whether student/TA conferences, where the collegiate student and TA converse, could further empower the students in their own education in the studio physics class format. The stakeholders for this project were primarily the students themselves, but also the departmental community composed of other faculty and TAs. The wider university and physics departments elsewhere can be seen as a larger community, although they were not directly involved in this study.


\section{Literature Review}
The authors engage with mentoring partnerships in a science classroom in conjunction with the principles of AR. This article primarily revolves around collegiate education, so it is helpful to define terms in this context. Because fewer studies exist on collegiate education, this literature review includes both collegiate and K-12 classrooms. For this AR study, the `student' is the college student, the `teacher' is a K-12 teacher, and the `faculty' refers to collegiate faculty members. The term `instructor' is used as an umbrella category, including teachers, faculty members, TAs, and undergraduate supplemental instructors (SIs). `Conference' is defined as a one-on-one meeting between an instructor and a student.

\subsection{Action research}
Although various definitions of AR exist, all revolve around the three key concepts of democratic participation, social justice, and community empowerment (\citeinpar{BM}; \citeinpar{DePalma}; \citeinpar{Somekh}; \citeinpar{Zeni}). In the context of AR, democratic participation refers to the practice of allowing the participants themselves to have a voice in the subject matter being studied, and in the progress of the study itself. Action researchers are no longer objective observers of their subjects, but instead all stake- holders work together to reach a common goal. Social justice is not merely allowing equal access to resources, because not all individuals are able to take advantage of supposedly equal access. To embrace social justice, one must be aware of past and current systematic inequities, and make changes to counteract these inequalities. The philosophy of community empowerment acknowledges that the researcher and participants exit with others. Indeed, community members can make valuable contributions to both the research project and the problem itself, with the goal of providing insight into the original research question(s).

\subsection{Action research in an educational setting}
AR is especially powerful in an educational setting, because K-12 teachers are traditionally powerless as the creators of pedagogy or the instigators of research (\citeinpar{IKP}; \citeinpar{Kayaoglu}). The movement of reflective practice or `teacher-as-researcher' does allow for both K-12 teachers and collegiate faculty to investigate their own teaching, but `students are often positioned as objects of teacher's study rather than collaborative partners or allies' (\citeinpar{BM}, 83). This study purposefully includes student stakeholders.

Beyond the AR practices described earlier, participatory AR as a form of reflective practice allows teachers to reflect upon their learning and influence the progress of an ongoing course (for example, \citeinpar{Anderson}). Moreover, the mandate of community engage- ment gives teachers a venue for soliciting feedback from colleagues, and has often been used in teacher training programs (for example, \citeinpar{Hoedebeck}; \citeinpar{Steele}). engaging students in their own education is one avenue to increase the democratic participation of students and reduce inequity. Simply allowing all students equal access to education and fair grading is not sufficient for social justice, as students have barriers to the access. Students from a lower socioeconomic class can ill afford educational materials ranging from pen and paper to textbooks and personal computers \citep{Tan}. In male-dominated fields such as physics, females may face (even unintentionally) additional social hurdles to their successranging from peer and parental pressure to study more gender-appropriate topics through to stereotype threat \citep{Hill}.

Reducing inequity in education requires that students are provided equal access to resources, and that proactive steps are made by those in power to assist students in accessing resources. Taking positive actions to reduce inequity allows students to reach a privileged state.

Community empowerment refers to allowing a larger set of stakeholders to voice their opinions and influence the course of action in the classroom. In the case of education, the community can refer to students' parents, the people of the town in which the school is situated, other educators, and academics or scientists at other institutions.

\subsection{Action research in the science classroom}
AR projects are common in science courses (for example, \citeinpar{Cavicci}; \citeinpar{NY}; \citeinpar{Nys}). These studies stress the importance of partnerships between learners and educators \citep{Cavicci}. They also highlight some difficulties of engaging in AR in a science classroom \citep{NY}.

Understanding one's own biases is important to the goals of AR (\citeinpar{NY}; \citeinpar{Nys}). Researchers must confront the assumptions inherent in their own viewpoints \citep{NY} - all researchers must struggle with overcoming this, but especially AR researchers due to their key position in multiple roles (e.g. educator, researcher). If AR researchers wish to fix problems of exclusion in classrooms, then they must first understand the existing power structure \citep{NY}.


\section{Partnerships}
The definition of `partnership' has evolved in the last decade, starting with vague partnership definitions and functions in educational settings \citep{Clifford}, and increasingly taking better form. \citet{B11} defines partnership, based loosely on \citeauthor{Goodlad}'s \citeyear{Goodlad} definition, as a planned, mutually beneficial, relationship between two or more parties which values differences; supports inquiry into practices to encourage change; expects dialog on expectations/goals/agendas; clearly outlines products and timelines; and, above all, builds relationships between participants by focusing on communication, trust, community issues, context, and culture. This definition grounds the partnership in the relationship, and people actually create the concept of partnership. Partnership itself is an abstract idea enacted by participants through interaction, comprised of people, and the relationship amongst the individuals, which is vital to the partnership's success (\citeinpar{B11}, \citeyear{B15}; \citeinpar{BH15}). All aspects of a school and culture interact as partnerships create products, such as physical objects, works of art, or written documents. The expectations, goals, and agendas of participants shape the partnership \citep{B11}. while partnerships may be most commonly thought of as being between two or more individuals of similar levels of experience and power, this is not necessarily the case with mentoring partnerships, such as a student and a teacher.

\subsection{Mentoring partnerships}
A mentoring partnership is defined as a situation where one person, a leader or mentor, acts as the catalyst for the partnership, increasing the rate of reaction. The mentor(s) in a part- nership focus on relationship-building and asking questions to enact changes where needed. The partnership group should decide on the areas of focus, and the individuals in a partnership ideally focus on each other's needs and strengths \citep{B11}. within the context of formal education, instructors most frequently take on this role of mentor. The notion of partner equality is important to partnerships \citep{LD}, and this becomes even more crucial in the context of an AR project: democratic participation of the learners necessitates that they are not the only ones receiving benefit, and social justice requires a neutralization of the power differential between mentors and mentees. This is similar to coaching in that both consist of two partners working towards a common goal with one partner having more experience. however, in coaching it is assumed that the coachee is the individual with a problem that must be solved, and thus will be the only individual benefiting (for example, \citeinpar{Gavriel}; \citeinpar{Passmore}).

\subsection{University student mentoring partnerships}
To a large extent, traditional classroom teaching is steeped in a power structure: the faculty has the knowledge, the faculty controls the classroom, and the faculty assigns the grade. This power is often expressed in language, such as the formality of speech \citep{Schoerning} or the assertiveness of the words and body language chosen \citep{Brown}. Similarly, the traditional modality of collegiate physics sets up the instructor as the ultimate authority in the classroom.

It is well known that any form of active learning (or student-centered classroom) results in better student outcomes than does traditional lecture or teacher-centered classroom pedagogy (\citeinpar{Beichner}; \citeinpar{Hake}). Studio physics begins to change the teacher-centered paradigm as the focus is shifted towards a learner-centered approach. When the Massachusetts Institute of Technology (with the Technology enabled Active learning project) switched to studio physics in the new format, `the instructor's ``command center,'' from which he or she can control all the technology in the room, was purposely placed in the middle to move the faculty member from a position of authority at the front of the classroom' (\citeinpar{Breslow}, 25).

The setting of studio physics is conducive to the breakdown of traditional lines of authority. 
AR's core value of democratic participation fits well with studio physics, making the course a good one on which to perform an AR project. 
Because of the lack of literature about the student/TA relationship and the call to examine the curriculum, the authors conducted this study about the student/teacher relationship, extending the concept of the curriculum to include the personal interactions.

As shown in the previous parts of this literature review, the authors propose that partnerships and mentoring are important to student success, but these strategies are often not used in traditional science classes. Conferences are another means to form a partnership with students (Goldstein and Conrad 1990) to solve issues that arise in the course \citep{Arbur}. Conferences show evidence of producing not only content gains but also improving student/teacher relationships; specifically, conferences 'facilitate a more personal and more ``real'', teacher-student interaction' (\citeinpar{Fassler}, 190). Furthermore, conferences allow for the negotiation of meaning \citep{Goldstein} rather than employing a top-down model from the faculty member. Unfortunately, there is a dearth of modern literature on the mentoring subject, and the older studies focus on English classes rather than STEM content areas.


\section{Theoretical framework and gap in the literature}
This AR study was framed using a social constructivist perspective where knowledge is built based on group interactions to produce meaning (\citeinpar{Vygotsky}; \citeinpar{Koro}). Thus, the authors viewed the data through a lens of community concept-building and understanding of the interactions taking place. This is similar to the approach of activity theory, situating the individual within his or her social context (for example, \citeinpar{McNicholl}), but differs in that the power of the individual is given precedence in accordance with the principles of AR. The goals of AR, social justice, democratic participation, and community empowerment touch upon a number of open fields of research relevant to this study, many of which are discussed in the Handbook of Research on Science Education \citep{Abell}. Further qualitative research of classroom environments is needed to under- stand the context of student learning \citep{Fraser}. while the role of the teacher in the K-12 science laboratory is well studied, the role of the TA in the collegiate science laboratory is not as explored \citep{Lunetta}. Physics classrooms are dominated by `teacher-centered instruction,' despite many findings that active student-centered learning is more effective, and approaches to changing this dynamic are needed (for example, \citeinpar{Hake}; \citeinpar{Duit}). There is also a lack of literature about student/ teacher conferences in science, about student/faculty conferences in higher education, and about student/TA conferences. As \citeauthor{DeBoer} states, `The challenge for all of us is to continue to think about how the curriculum can be used to improve the teaching and learning in science for all' (\citeyear{DeBoer}, 576). Student/TA conferences are a means of manipulating the standard curriculum for student opinions and concerns to be expressed through instructor/student interactions.


\section{Research questions}
The authors sought how to utilize student/TA conferences to address aspects of partnering, mentoring, and inequity that occur in the collegiate classroom. The main goals of the study were to reduce inequity and provide college science students with a voice where usually that voice is silenced. These questions are written as they were asked ? that is, in the first person ? and include the following: 
(1) `How difficult will it be for us to empower the students to use their voice for personal success in the classroom?'; 
(2) `What sorts of successes and barriers exist for giving the students a voice in certain situations, and are these successes/ barriers within the students themselves, within us as educator-researchers, or within the other community members?'; 
(3) `Will we be able to establish a rapport between students and one or more mentors in science?'; 
(4) `Will we see any evidence that establishing a rapport actually helps to reduce inequity for our students either in the classroom or in their greater lives?'; 
(5) `Will having student/TA conferences allow me (the TA) to better learn and remem- ber students? names to enable a more personable connection?'; and, finally, 
(6) `How will this process unfold and to what extent will our efforts change the culture in the traditional science classroom?' 
These questions can be organized into three major themes, as follows:

\begin{enumerate}[I.]
\item Establishing a rapport with a mentor (Questions 3, 4, and 5).
\item Empowering students and reducing inequity (Questions 1, 4, and 5).
\item Successes and barriers to hearing students' voices (Questions 2, 5, and 6).
\end{enumerate}


\section{Methodology}
The authors used the three core principles of AR to address issues of systemic inequality in the classroom, and they attempted to give students a voice to improve their democratic participation. The authors had numerous conversations with other members of their academic communities, and they focused on reducing known issues of social injustice. In addition to this AR focus, both quantitative and qualitative data were collected to determine whether these goals were being met. This section is organized with a description of the sample, the student/TA conference format, and specific methods of data collection. The methods are further broken down by viewpoint and goal of the particular form of data collected.


\section{Sample}
All 36 students in the studio physics course were invited to participate in the study. Twenty- nine students (81\%) chose to participate, and 27 (75\% of the class or 93\% of participants) attended conferences. To reduce the chance of triggering stereotype threat, the demographics for this study were based on the TA's observations of student gender and race/ethnicity. The demographics consisted of 22 (76\%) white males, six (21\%) white females, and one (3\%) hispanic male.


\section{Methods and student/TA conference format}
This study took place over the course of nine weeks during the Fall of 2014, and the data collection methods were various and included: interview, survey, observational field notes, and personal reflections. More detail on these methods follows. The original goal of the data collection was to determine the empowerment of the students within the class, and data were collected to identify possible barriers in the participants' self-expression. The student/TA conferences consisted of 15-minute informal conversations between a student and the TA. The purpose of the conferences was for the TA to learn more about the whole student, such as their interests, families, reasons for taking the course, academic goals, and learning style, and was intended to give the participants an outlet for their voice to be heard and respected. It was not expected that all of these topics would be made clear in every conference due to the short duration, but these were the types of topics that the TA touched upon with the students. The questions asked (Appendix A, see Supplemental Material) served only as a starting point for conversation, rather than as strict guidelines.

The following subsections further detail the methods of the particular forms of data collected.

\subsection{Viewpoint of participants and community}
The viewpoints of the participants were explicitly solicited during each 15-minute one- on-one conference (27 times). The post-survey included both specific questions soliciting student input and an open-ended question prompting students to share additional insights about the study or their participation in it. The TA expressed that she was open to hearing from the students at numerous other instances, including class time, office hours, by appointment at other times, or via email.

The TA also held informal conversations about the course, teaching, and student learning in general, with the studio physics faculty, the two non-studio physics faculty, the TAs for the non-studio sections, and the SI for one of the non-studio sections. These conversations took place in person, either by the TA dropping in to the office of the faculty members or by running into the other TA or the SI in the hallway. The TA reflected upon her interactions with the students and these other educators throughout the semester, and after the termination of the study.

\subsection{Qualitative data collected}
The pre-survey (see Appendix A, see Supplemental Material) was administered during class time before the commencement of student/TA conferences and after their first examination, and contained both quantitative (yes/no, ordinal, and Likert, as per \citeinpar{Likert}) and qualitative (free response) questions. The post-survey included five additional free response questions not on the pre-survey prompting participants to address other ways the TA could have helped them to feel more comfortable asking questions during class, outside of class, or to help the student's learning and motivation. The final question asked whether participants wanted to share anything regarding the study or their participation in it. The TA kept observational field notes during student/TA conferences and during class time (total of eight pages), and wrote additional reflections both during and after the course sessions (total of 17 pages).

\subsection{Quantitative data collected}
Author 1, classically trained in the pure sciences but stretching into AR, valued quantitative data, and thus felt reassured using a survey as part of the data-set. There were nine questions that were answered on a bubble sheet. Counts of student/instructor interactions were recorded on eight class dates (approximately three weeks of class, or 16 contact hours) after the completion of the student/TA conferences. Two assistants were recruited to count student/instructor interactions in non-studio sections of the course for comparison with the studio section. The post-survey was administered in class after the completion of the tallies.


\section{Analysis}
The authors analyzed the data including field notes, survey responses, and tally of interactions in a number of ways, including searching for the voice of the participants and community, qualitative coding for themes, and quantitative statistical calculations. These approaches are shown in the following sections.

\subsection{Voice of the participants and community}
The authors sought commonalities and patterns within the data throughout the analysis. Some of the patterns were discovered in participant answers to free response questions on pre/post-surveys, and in conversations with the authors and other community members involved in the study. Input from community members was gained through emailed and face-to-face discussions. In addition, the authors reflected upon the process and their experiences both individually and in correspondence with each other.

\subsection{Qualitative}
The authors coded participants' answers on free response questions for themes, and looked for individual words or phrases that one participant repeated multiple times, or that multiple participants used. when ideas were found to recur across different participants within the study, the responses of other participants were examined for similar ideas using different words. These ideas were then grouped into overarching ideas.

\subsection{Quantitative}
The authors recorded the number and percent of yes/no question answers, as well as the number of `Agree' or `Strongly Agree' responses to Likert-scale questions. The authors compared the counts of student/TA interactions between different sections of physics (studio and non-studio) and between different class styles (lecture, discussion, and lab). For the studio physics section, `lab' was defined as those times when the lab manual was in use. `Discussion' was defined as the times when the students were working in groups on work- sheets, primarily addressing conceptual questions. The `lecture-like' modality included teacher-centric activities where the instructor talked while displaying lecture notes on an lCD projector, demonstrated problem-solving, or had students respond to clicker questions. These numbers were analyzed using statistical methods after the course for triangulation with the non-quantitative data.


\section{Results and findings}
The following results and findings are separated into sections on the three themes that emerged from the research questions: establishing a rapport with a mentor; empowering students and reducing inequity; and successes and barriers to hearing students' voices. Table~\ref{tab:Survey} showcases selected student responses on the post-survey, broken down by category.


\begin{table}[!htb]
\begin{center}
\begin{tabular}{p{4.5cm} p{5.5cm} p{5.5cm}}
\hline \hline
Theme & Successes & Barriers \\
\hline 
I: Establishing a rapport with a mentor &
In response to how the TA could motivate students to talk about their personal issues, `Express her personal issues.'

`Offer points back on assignments if we come discuss it in office hours.' & 
`i think it is just a personal problem that I would need to focus more and do more work outside of class.'

`Not really one to talk about personal problems with people i don?t know very well.'

`Teachers/professors shouldn't be too involved in personal life.' \\
\\
II: Empowering students and reducing inequity & 
`Not grade as harshly on assignments, labs, and tests when we clearly have the concepts and only messed up the work. Easier to want to do better when we're not afraid of being thrown on the chopping block for small mistakes.'

`Maybe encourage students to set up study groups.'

`Post a \% grade! It bothers me not knowing, and if I had a lower grade I would work much harder.' &
`Not interrupting the teacher while he is teaching to interject her opinion on how to do a problem.'

Many responses were about clarifying or expanding the existing power structure (e.g. `clearer guidelines for lab reports/ homework,' `hold study sessions every week.') \\
\\
III: Hearing student voices &
`I think that more application examples could be used in class like where these principles in class are applied.'

`This study seems bias[ed] due to only yes/no answers on opinion, but could definitely be applicable.'

`I don't believe this study had anything to do with motivation. It only helped you be more open to questions.' &
Four of the six free response questions had more blank and non-substantive responses (e.g., `Fine as is, ``nothing") than substantive responses.

A total of 61.0\% of free-response questions were either left blank or had non-substantive responses. \\
\hline
\end{tabular}
\end{center}
\caption{Qualitative data showcasing selected post-survey responses.}
\label{tab:Survey}
\end{table}

\subsection{Theme I: Establishing a rapport with a mentor}
The TA (Author 1) has extensive experience interacting with students as a community college faculty, but not in a larger university setting. During the physics conferences, some students were comfortable just chatting while others needed additional prompting. Students asked questions about the TA as well (e.g. hometown) and this sometimes led to conversations about the benefits and drawbacks of both large urban areas and smaller, more close-knit communities. Many of the students expressed their appreciation of the outdoors.

A few students preferred to use the time for tutoring on the course content, and the TA followed the students' guidance rather than forcing the students out of their comfort zone. 
Some students preferred to discuss specific problems from their homework assignments. while this choice may not have directly fostered the emotional connection of a mentorship, these students received assistance they might not have sought out otherwise.

On the pre/post-survey, the second and third questions dug deeper into the perception of how well the TA knew the students. On the pre-survey 12.5\% of respondents felt that she knew them as students, while on the post-survey 34.3\% said they felt she knew them as students. On the pre-survey 0\% said they felt that the TA knew them as individuals, while on the post-survey 34.3\% responded that she did. The additional Questions 10-12 on the post-survey revolved around the students' comfort with the TA. eighty percent of respondents said that they felt comfortable asking the TA questions during class, while 69\% felt comfortable doing so outside of class.

Only one individual (3\%) reported feeling comfortable talking to the TA about personal issues or for emotional support. Seven students had substantive responses to the free response post-survey question about sharing personal issues with or seeking emotional support from the TA (see examples in Table~\ref{tab:Survey}.). Three students said they would never bring personal issues to a TA (such as Author 1) or professor, and two students said they would only bring such issues to people with whom they felt close.

The final three multiple-choice questions on the post-survey were about student perceptions on knowing the TA. All participants indicated that they knew the TA's name. Students were evenly split on whether they felt that they knew the TA's strengths and weaknesses as an educator (43\% yes, 46\% no, 11\% no response). The minority of students felt that they knew the TA as a person, including her hobbies or interests (63\% no, 23\% yes, 14\% no response).

\subsection{Theme II: Empowering students and reducing inequity}
The authors addressed inequity through student voice in this AR study. The participants frequently referred to external motivators for the course on the post-survey. Grades served as a strong motivation, and a few participants suggested that providing extra credit would give them more motivation. Increased grade feedback was suggested as another motivator, such as posting their grades to either the school-provided learning management system or to the publisher-provided homework system. Five students referred to internal motivation, mostly saying that the TA or professor cannot motivate the student, but that students can motivate themselves. See Table~\ref{tab:Survey} for examples of these comments.

Students expressed reluctance to break the hierarchy of faculty being `superior' to TAs, who are in turn `superior' to students. Table~\ref{tab:Survey} highlights two instances of this: in one case the student chastised the TA for breaking the hierarchy, while in the other the student asks the TA to intervene on the behalf of students.

\subsection{Theme III: Successes and barriers to hearing students' voices}
\subsubsection{Successes and barriers within the students themselves}
One student discussed how his motivation to succeed was based upon the fact that he was on a full scholarship with a stipend, and had an infant child and a wife to support - should he lose the stipend, he would be unable to support his family. Another was so wrapped up in the conversation that she talked for more than 20 minutes, five minutes longer than the intended duration of the conference. One student talked about his ambition of becoming an audio engineer, which the TA referenced when the topic of tuning circuits consisting of a resistor, inductor, and capacitor (RlC circuits) emerged in class.

Many of the responses on the post-survey were opinions on how the course could be improved or how the instructors could give additional feedback to the students. On the other hand, students were often reluctant to speak up and voice their opinions aloud. This was also reflected in the responses on the participants' post-surveys, where many students responded with rote phrases such as `fine as is,' `things are fine,' or `nothing,' as shown in Table~\ref{tab:Survey}.

\subsubsection{Case study: Brandon - successes and barriers}
One student in particular, who is referred to as Brandon, is an interesting case and the authors explore his situation closely here. he was struggling in physics and had already been contacted by the TA for his poor attendance. 
Because only three students in the course had poor attendance during the semester, Brandon's absenteeism was conspicuous. 
Brandon failed to show up for his scheduled conference and the TA approached him about this when she next saw him. he expressed contrition and they scheduled another meeting one week later. 
At this conference, Brandon volunteered contributing factors to his absenteeism, and he expressed concern at his lack of progress in the course.

The TA and Brandon decided to meet every week for the final six weeks of the semester. Brandon attended approximately half of these sessions, without informing the TA beforehand of intended absences. Brandon brought the work he had begun on that week's homework assignment to these meetings, and the TA assisted him to complete as many problems as possible over the course of an hour. This led Brandon to complete more of the homework assignments than he would have had he not met with the TA.

The course professor also reported that Brandon was meeting with him occasionally (with even less regularity) to go over material. Brandon had disclosed more of the circumstances of his absenteeism to the faculty, and the faculty ended up introducing Brandon to another graduate student with similar life circumstances who could serve as a mentor in a similar situation. 
Brandon's significance to this study is explored later in the article.

\subsubsection{Successes and barriers within the educator-researchers}
At the start of the study, the TA knew approximately 50\% of students' names. After the student/TA conferences, she recalled 77\% of the names. During the last week of the semester, the TA found that she could recall 75\% of students' names from their faces. Thus, the TA had an increase in name recognition and personal student connections.

\subsubsection{Successes and barriers within the university physics community}
In addition to talking with the student participants in the student/TA conferences and during class time, she also interacted with the other course instructors. 
The SI of one non-studio physics section was glad to help with this project. he was fascinated with the process of education research, and eagerly suggested ideas to attempt. 
In fact, it was the SI's suggestion to separate the counts of interactions by what style of teaching was happening at the moment (lecture, discussion, or lab-like), and the authors embraced and incorporated the idea into the AR project.

Author 1 reached out to all four graduate TAs for the non-studio sections of the course, and found one TA willing to help with this study. The professor leading the studio physics section expressed an interest in doing student/faculty conferences himself, but he conducted none during his class. From observing the class and from previous interactions, the professor appeared vested in student learning, but did not take the next step into personal student conferences.

Another professor, known to Author 1 for caring about student learning, ascribed to a positivist or teacher-centered model of education. Author 1 and this professor had previously held a number of conversations where he expressed distrust in the results of educational research. he attributed the success of learner-centered teaching methods to a reaction effect where students respond positively to any change in their learning environment. The conversations between Author 1 and this instructor about this project were more frequent than with other faculty members. Based on these interactions, the authors attempted to capture the voices of the instructors for this AR project.

\subsubsection{Successes and barriers of the studio physics format}
To provide context, a table of data (Table~\ref{tab:Interactions}) is presented for the reader. This table includes the: average number of students enrolled in each section; the number of sessions observed; the number of interactions; the calculated interaction rate (i.e. the number of interactions per session per student); and the ratio of this value for studio physics to non-studio physics. The interaction rate is expressed as the number of interactions per student per session, and thus is scaled by the number of days observed and the size of the class. larger numerical values show more interaction. In lecture sessions, studio physics students interacted 2.7 times more with their TA than non-studio physics students did with either of their instructors. The values for discussion sections are similar between the two forms of physics.


\begin{table}[!htb]
\begin{center}
\begin{tabular}{l c c c}
\hline \hline
   & \multicolumn{3}{c}{Modality} \\
   & Lecture & Discussion & Lab \\
\hline 
\bf{Studio physics (after conferences)} & & & \\
  Number of students enrolled  & 36 & 36 & 36 \\
  Number of sessions observed  & 8 & 7 & 5 \\
  Number of interactions  & 37 & 72 & 60 \\
     Interaction rate & 0.128 & 0.286 & 0.333 \\
\\
\bf{Non-studio physics (no conferences)} & & & \\
  Number of students enrolled (average per section)  & 63 & 35 & 26 \\
  Number of sessions observed & 4 & 2 & 2 \\
  Number of interactions & 12 & 15 & 28 \\
     Interaction rate & 0.0476 & 0.214 & 0.538 \\
\\
\bf{Ratio of studio/non-studio interactions rate} & 2.7 & 1.3 & 0.62 \\
\hline
\end{tabular}
\end{center}
\caption{interaction rate (number of interactions per session per student) in the different modalities.}
\label{tab:Interactions}
\end{table}


\section{Discussion and conclusions}
The students participating in this study showed evidence of enjoying the student/TA conference process. They talked about themselves and asked questions about the TA as a person. however, it was difficult to engage students as active partners in either their education or the study itself. even those participants who enjoyed the conferences were reluctant to voice their opinions during or after class time. while the authors did not explicitly explore the causes of this reluctance, the pre-existing power structure of the classroom - the traditional teacher in a seemingly superior role - is speculated as a contributing factor.

This section discusses the first two themes of establishing a rapport with a mentor (including Brandon's case), and empowering students and ameliorating inequity, a modified third theme of improving opportunities for hearing students' voices, and also explores limitations of the study.

\subsection{Theme I: Establishing a rapport with a mentor}
Rapport is built upon interactions, and the TA observed that it was harder to develop a rapport with students at the research university than at a community college. 
Factors such as larger class size and fewer student contact hours per instructor at the research university could be part of the explanation. 
However, the interactions were improved when she first served as a TA in studio physics the semester before this study - the smaller class size and larger number of contact hours in studio physics allowed the TA to get to know her students, but there was room for growth. 
Overall, the TA engaged the students more with conferences, and built a mentoring rapport with Brandon, but neither that connection nor any others persisted forwards. 
In the future, the TA should create a follow-up plan to keep lines of communication open.

\subsubsection{Case study: Brandon - discussion}
The student/TA conferences had a beneficial impact for intervention in cases of poor student performance. 
Traditionally, the initiative must come from the student for performance to change. 
If an instructor only schedules meetings with struggling students, then students may not possess any initiative to improve their work, and negative connotations of student/teacher meetings may dominate the interaction. 
On the other hand, if all students are meeting with the instructor, this could provide the extra push for the student to voice an opinion.

In Brandon's case in particular, there was an improvement upon his original situation. 
The student's efforts to sometimes attend weekly meetings with the TA may be evidence that his motivation increased. 
While another student mentioned motivation and the TA's role by stating `[letting] me know at a personal level that my education and future is important to her,' it remains unclear whether the conference for this particular student had a positive effect on motivation.

\subsection{Theme II: Empowering students and ameliorating inequity}
\subsubsection{Empowering students}
This study empowered collegiate studio physics students in their education, but found challenges to this AR process, primarily from the students themselves. 
Many of the students did exhibit enjoyment and eagerness to discuss personal stories with the TA. 
However, they were considerably more inhibited when it came to expressing opinions about the course or the study, whether face to face or on paper.

The sheer difficulty of soliciting feedback from the students was surprising to the TA, whether in person or on the post-survey. 
Many of the students responded with rote responses such as `Fine as is,' borrowing the wording suggested on the survey. 
A number of other students responded that discussing personal matters with the TA was either not appropriate or not something they would normally do. 
Perhaps this AR study was such an anomaly to the physics students that they were unable to balance the seeming power differential and their solicited voice.

Students exhibited external motivation on their post-surveys, such as discussing extra credit or food as motivators. 
However, their wish for more grading feedback could either support this wish for more external motivation (grades) or could imply that they are looking for assistance with their metacognitive processes. 
One student (appropriately) criticized this study: `I don't believe this study had anything to do with motivation. It only helped you be more open to questions.' 
Possibly, this student, although rightfully cautious and skeptical, took the first steps towards finding and using a voice with this strong statement.

\subsection{Ameliorating inequity}
While the authors kept best intentions for ameliorating inequity in their classrooms, providing the students a voice in the educational process was the only means of addressing injustice. 
The students felt that the TA cared about their opinions of the course, yet they were reluctant to vocalize substantial feedback. 
There are multiple possible causes for this lack of student expression. It could be a holdover of the power structure in which the TAs are superior to the students. 
A TA's status in the classroom hierarchy (i.e. `below' the professor) might have led the students to feel that voicing their concerns would negatively travel to the professor or `one in charge.' 
The students may have felt that the class was optimally effective for them, although the TA feels this is unlikely because there were students who did not pass the course. 
Lastly, it could be that the students' metacognition or self-awareness is weak, or they felt uncomfortable when the TA asked for useful feedback.

\subsubsection{Student names}
The students' perceptions of the TA's knowledge of student names appeared to be significantly different from her own perception of how well she knew their names. 
Over the course of the study, the students did feel that the TA improved in how well she knew the students, both with regards to their strengths and their weaknesses as students, and as individuals. 
The TA's perception of her knowledge of student names also improved over this AR study. 
The students' answers to the questions about how well the TA knew them do show evidence that her personal connection to the students increased. 
Learning student names is an important step towards recognizing students as complete human beings, thus giving them more power and validating their role as active contributors to the classroom experience.

\subsection{Modified Theme III: Improving opportunities for hearing students' voices}
The authors identified a number of future opportunities to help students voice their ideas, most obviously to explicitly encourage and value student participation and openness. 
However, one possibility for the students' reluctance to voice an opinion is that they could have unused metacognition skills and may not understand how to assess what would enhance their learning. 
Another possibility is that the power structure of the classroom could be stifling those students who do have ideas for feedback. 
Students could be afraid that expressing their opinions could be taken as criticism of the TA or the instructor in charge of the course, and they might fear inadvertently negatively impacting their grade. 
A third possible cause is that students could just be accustomed to the social contract of sitting quietly in class and parroting back what the instructor says on examinations, rather than actively engaging with their learning and even improving it.

Author 1's own reluctance to press the students for more information was an unexpected barrier, and is one that she needs to examine critically. There is a boundary of when it is acceptable to lead students past their comfort zone for the sake of their metacognitive growth. As an experienced physics teacher, the TA knows where this boundary lies with regards to academic content and skills growth, but as a novice AR researcher she does not yet know where it lies with regards to student empowerment. As a result, when the students seemed uncomfortable, she preferred to err on the side of light pressure. Author 1 needs to grow as an AR researcher to learn where the acceptable boundaries lie.

Another issue was that there were difficulties in working with the other TAs and faculty for the non-studio course. Some of these were partially due to departmental cultural customs regarding organization of TAs and faculty for large courses. The physics department has a limited number of mechanisms in place for interaction of instructors. On weeks when faculty meetings did not take place, one of the non-research faculty members typically scheduled `teaching chats,' but these were limited in scope. The organizing individual was faculty for one of the two non-studio physics sections, specifically the one who has expressed skepticism of educational research. For the semester in question only one teaching chat was scheduled, and it involved a demonstration of display technology rather than actual discussion of the teaching or learning in the courses. In addition, attendance at these teaching chats was low, typically in the single digits. This small group has the potential to begin change with student voices, but the group has not yet embraced this culture.

In future semesters, Author 1 wishes to reach out further to other members of her department. Many physics professors are skeptical of the results of educational research, thinking that while more active forms of teaching may work in certain situations, they will not work in their own situation. At research universities, departments are focused on content rather than on teaching pedagogy and the importance of student interactions. Thus, it is often an uphill battle to convince instructors to adopt the results of educational research. The authors continue to encourage faculty to perform their own studies, empowering them to make their own investigations, and inspiring them to involve their students' opinions.

\subsubsection{Number of student/instructor interactions}
The authors found it surprising that the interaction rates for lab sections were greater in non-studio physics than in studio physics. There are a number of factors that could contribute to this, including the three described here. First, the studio physics numbers referred only to the number of interactions the students had with their TA. If the students had the same number of interactions with the faculty, then the interaction rate might be doubled to 0.667 - greater than the non-studio physics value of 0.538. Secondly, if non-studio physics students ask fewer questions in lecture, perhaps they need to ask more questions of their TA than the studio physics students do. lastly, the introduction to the studio physics labs was always given by the faculty, so it was possible that the professor explained the lab more clearly than did the TA of the non-studio labs. Despite these suggestions, data can be interpreted as evidence that the TA communicated with her studio physics students post-conference more than did other instructors with their non-studio physics students.

\subsection{Limitations}
There are several limitations to this AR study. These can be separated into categories of the sample, the study design, and selection effects and demographics of the assistants and TA. These limitations are now discussed.

The sample size and the lack of diversity in the sample result in limitations to the study. The small sample size means that it cannot be generalized quantitatively to larger populations, only qualitatively. no barriers to either women or non-white domestic students were observed, perhaps due to the student demographics: 22 (76\%) white male, six (21\%) white female, and one (3\%) Hispanic male.

The study design has a number of limitations. Qualitative data were collected and analyzed by a single individual, so no inter-rater reliabilities are possible. The interaction score was normalized as per session, but sessions were of different lengths depending upon the modality - all studio physics class meetings were one hour and 50 minutes in length, while non-studio physics lectures studied here were one hour and 15 minutes, discussions were 50 minutes, and labs were one hour and 50 minutes. Another non-studio physics lecture section with 50-minute sessions was not studied here. It is possible that the lack of complete anonymity on the post-survey may have contributed to this as participants were asked for their names to track changes. names were used rather than codes primarily so that the TA could make connections between discussions with the students and their responses.

The selection and demographics of the TA herself and her assistants are another limitation. 
Only one TA from a non-studio physics section was willing to assist with this study, so there may be selection effects as a result of only comparing that TA's sections with the studio physics class. 
The TA for the studio physics class (Author 1) was female, while all faculty and assistants (one TA and one SI) in both the studio physics class and the non-studio classes were male, which could have introduced additional power differential effects.

\section{Implications}
educators are often searching for ways to test new potential solutions for problems in the classroom. AR not only addresses this need, but is also a radical approach to science education (especially in physics) in which some power is moved from the hands of the educator to the hands of the students and community. when looking to form a mentoring partnership between students and educators in science classes, AR can be a powerful tool for reducing inequities and engaging students. This study provides a path for novice AR researchers and helps to identify some of the successes and challenges inherent in taking on a beginning AR project.

\subsection{Mentoring}
The mentoring partnership is one that is inherently unbalanced, with one individual taking an experienced role and the other being a novice learner. 
It can be difficult to overcome this inequity, and the learning partner may resist efforts to overturn this system. 
When a mentee seeks out a mentor precisely to learn from her/his expertise, this sets up a power differential that the mentee may feel is inherent to the learning. 
Similarly, the mentor may feel that she/ he deserves to be the person of greater power in the relationship due to her/his greater knowledge. 
But without the need of the mentee to learn, the mentor's role would be empty, and therefore a mentor should take care to elicit the mentee's needs within the mentoring partnership.

Educational research increasingly acknowledges the necessity for teachers to meet students at their level, for example using differentiated instruction. 
Moreover, the very philosophy of AR in the context of education requires that the educator-researcher determines the systematic inequities standing in the way of each student's access to education. 
An educator-researcher who does not determine her students' challenges cannot meet her students' needs, and therefore will not be espousing the philosophy of social justice.

\subsection{Partnerships}
While partnerships are important in STEM disciplines, the interpersonal skills learned carry over into many additional fields and careers. Communication skills learned through partnerships serve as a stepping-stone for all career paths, and can help improve the workforce. Students who learn to establish and work in a partnership in school can transfer this skill to working with supervisors, peers, and supervisees in their professional career.

\subsection{Action research}
In a phrase so often quoted that it has been attributed to many sources, `social justice is defined as promoting a just society by challenging injustice and valuing diversity.' 
In its ideal form, AR involves the participants and their community working with the researcher(s). 
Together all stakeholders identify the problems they face, and pursue solutions which allow individuals and their societies to flourish together. while the researcher can serve as a guide in this process, her voice is no more significant or important than that of any other person involved. AR allows oppressed peoples to address issues that concern them and participate as fellow researchers. Importantly, the ultimate goals of AR in education are twofold: to empower students in their education so that they are equally responsible for their learning and the classroom environment; and to bring them into the research process so that they are equally accountable for the creation of new knowledge in educational research with strong voices.


\section{Acknowledgement}
The authors would like to thank the participants of the study. Andria C. Schwortz would like to thank the members of the University of Wyoming Physics Department for their support during the course of this study.

The Version of Record of this manuscript has been published and is available in Educational Action Research, 03 October 2016, http://www.tandfonline.com/, DOI:10.1080/09650792.2016.1221838. 

\section{Disclosure statement}
No potential conflict of interest was reported by the authors.

\section{Funding}
The authors acknowledge partial funding for this study from the grants LASSI [DOE WDE MSP \#WY140202]; and SQARMS [NSF DUE \#1339853].


\bibliographystyle{apj}
\bibliography{EAR2016}


\end{document}